\documentstyle[epsf,12pt]{elsart}
\begin{document}
\begin{frontmatter}
\title{\large TIME CORRELATIONS OF HIGH ENERGY MUONS IN AN UNDERGROUND 
DETECTOR}

\author[bologna]{Y. Becherini},
\author[bologna,tesre]{S. Cecchini},
\author[bologna]{T. Chiarusi},
\author[bologna]{M. Cozzi},
\author[oujda]{H. Dekhissi\thanksref{corr}},
\author[oujda]{J. Derkaoui},
\author[bologna]{L.S. Esposito},
\author[bologna]{G. Giacomelli},
\author[bologna]{M. Giorgini},
\author[bari]{N. Giglietto},
\author[oujda]{F. Maaroufi},
\author[bologna]{G. Mandrioli},
\author[bologna]{A. Margiotta},
\author[bologna,pin]{S. Manzoor},
\author[oujda]{A. Moussa},
\author[bologna]{L. Patrizii},
\author[bologna,ro]{V. Popa},
\author[bologna]{M. Sioli},
\author[bologna]{G. Sirri},
\author[bologna]{M. Spurio} and
\author[bologna]{V. Togo}
\address[bologna]{Dipartimento di Fisica dell'Universit\`a di Bologna and INFN
40127 Bologna, Italy.}
\address[tesre]{IASF/CNR, Sez. Bologna, 40127 Bologna, Italy.}
\address[oujda]{LPTP, Faculty of Sciences, University Mohamed Ist, B.P.424,
Oujda, Morocco.}
\address[bari]{Dipartimento Interateneo di Fisica del 
Politecnico-Universit\`a di Bari and INFN, 70126 Bari, Italy}
\address[pin]{PRD, PINSTECH,P.O. Nilore, Islamabad, Pakistan.}
\address[ro]{ISS, 77125 Bucharest-Magurele, Romania.}
\thanks[corr]{Corresponding author. Tel: 212-56500601; Fax: 212-56500603.\\
{\it E-mail address}: dekhissi@sciences.univ-oujda.ac.ma (H. Dekhissi)}
\begin{center}Accepted for publication in  Astroparticle Physics.
\end{center}
\begin{abstract}
We present the result of a search for correlations in the arrival times
of high energy muons collected from 1995 till 2000 with the
streamer tube system of the complete MACRO detector at the underground
Gran Sasso Lab.
Large samples of single muons (8.6 million), double muons (0.46 million)
and multiple muons with multiplicities from 3 to 6 (0.08 million) were
selected. These samples were used to search for time correlations of cosmic
ray particles coming from the whole upper hemisphere or from selected space
cones.
The results of our analyses confirm with high statistics a random arrival time
distribution of high energy cosmic rays.
\end{abstract}
\end{frontmatter}

\section{Introduction}

It is generally expected that high energy (HE) galactic cosmic rays (CRs)
have a random arrival time; but it was also suggested that CRs coming from a
nearby pulsar, a nearby supernova-like event, etc. could show modulation
effects \cite{Weekes}. Some early experiments reported non-random components
in the arrival times of HE cosmic rays \cite{Bhath}, but higher precision measurements 
did not see any effect, see e.g. ref. \cite{Ahlen1}.\par

MACRO was a large multipurpose detector installed in hall B of the Gran
Sasso National Laboratories (LNGS) at the minimum depth of overburden rock of 3200 mwe
(average depth of 3600 mwe). The
detector had a modular structure consisting of six supermodules (SMs) each of
dimensions 12 m x 12 m x 9.3 m. It was made of three horizontal layers of
liquid scintillation counters, 14 horizontal layers of streamer tubes, one
horizontal layer of nuclear track detectors and seven layers of rock absorbers.
The vertical sides were closed by six layers of streamer tubes and
one layer of scintillators. The complete detector was thus a nearly "closed box" with
a total length of 76.7 m \cite{Ahlen2}; it had a lower part
4.8 m high, with 10 horizontal layers of streamer tubes, two layers of
scintillators and seven layers of rock absorber. The upper part ("attico")
contained also the electronics.
The main goals of the experiment were the search for rare particles,
like magnetic monopoles and nuclearites \cite{Ambrosio1}, the study of
atmospheric neutrino oscillations \cite{Ambrosio2} and the study of downgoing
HE muons \cite{Ambrosio3}\cite{Ambrosio3-1}. \par
In this paper we present a high statistics study of the time distributions
of downgoing HE single, double and multiple muons measured from 1995 till 2000 with
the streamer tube system. The investigation
concerns also the time distributions of muons coming from selected directions in the sky.\par
For each muon arriving at time t$_0$, we studied the
distribution of the time intervals elapsed from $t_0$ till the arrival times $t_1$, 
$t_2$, $t_3$, $t_4$, $t_5$ of the next five muons: (t$_1$-t$_0$), (t$_2$-t$_0$), 
(t$_3$-t$_0$), (t$_4$-t$_0$) and (t$_5$-t$_0$). The present statistics is larger 
by at least a factor of 20 than those of previous underground experiments 
\cite{Bhath}\cite{Ahlen1}.\par
The data selection is discussed in Section 2, the time correlations
in Section 3 and the conclusions in Section 4.

\section{Data selection and tests analyses}
For this analysis we selected runs according to the following criteria:\\
a) Run duration greater than two hours, (the average run duration was about 5 hours).\\
b) Muon rates R in the range 840$\le$R$\le$960 muons per hour (corresponding to deviations
of less than $2\sigma$ from the average).\\
c) Acquisition dead time smaller than 0.4\% for each of the 3 microvax computers.\\ 
d) No errors in the atomic clock (The arrival time of muons was
measured with an atomic clock with a precision of about $1\mu$s absolute
and 0.5$\mu$s relative).\\
e) No trigger problems.\\
f) No problems with the streamer tube gas system.\\
g) Streamer tube efficiencies of the wires and strips larger than 90\% and
86\%, respectively (the average wire efficiencies for each selected run is
about 94\%).\par
Further selection criteria:\\
A single muon event should have a track with
multiplicity equal to one, both in the wire and strip streamer tube views;
double muon events must have 2 tracks in the wire and strip views and multiple muons
have multiplicities from 3 to 6 and reconstructed on both
wire and strip views.\par
2035 runs passed the cuts for a total number of 8600$\times 10^3$
 single muons, 460$\times 10^3$ double muons and 80$\times 10^3$ multiple muons.\par 
The primary cosmic ray energy
for events with a single muon is larger than 20 TeV. For events
with two (multiple) muons the mean primary energy is larger than 200 TeV (1500 TeV).\par 
\begin{figure}[h]
\begin{center}
   \mbox{\epsfysize=12cm
         \epsffile{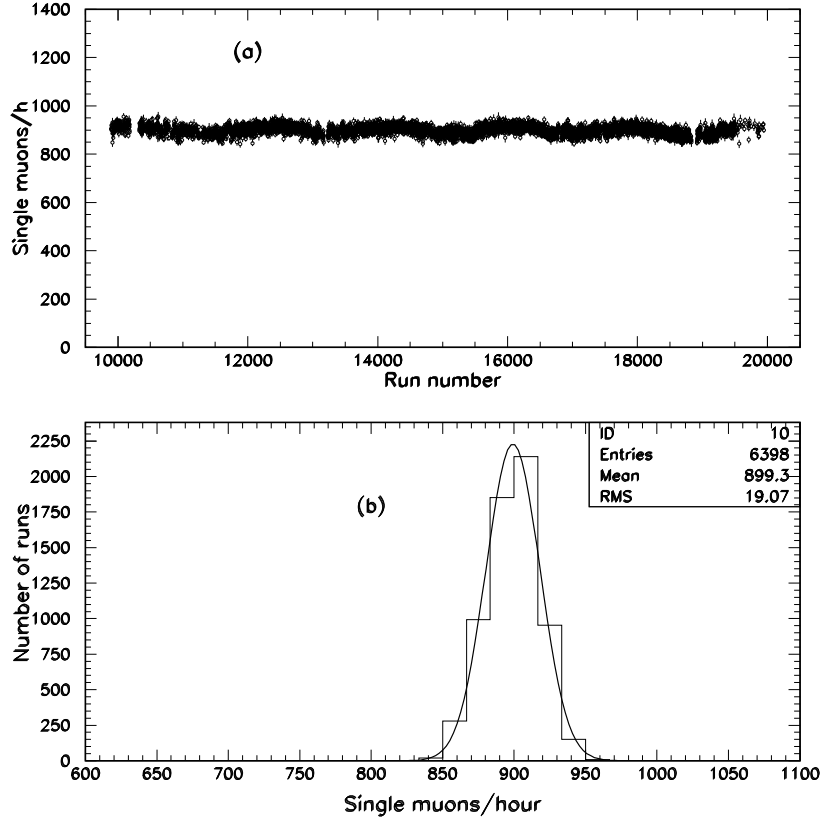}}
\end{center}
\caption{(a) Single muon rates per hour for selected runs. The observed rate variation 
is the signal of the seasonal effect.
(b) Distribution of the hourly counting rate per run. The average rate was 899 muons/hour.} 
\label{fig:mrate}
\end{figure}

The distribution of single muon rates per hour (taken when the full apparatus was 
running) is shown in Fig. \ref{fig:mrate}a. The average run duration was about 5 hours.
Fig. \ref{fig:mrate}b shows the number of runs versus the muon rate of each run. For the whole 
apparatus the average rate was 899 muons/hour.   

In Fig. \ref{fig:mrate}a we notice that the counting rate per run shows a regular 
variation with run number (time). This is the signature of the seasonal variation 
of the muon flux 
due to density and temperature changes of the upper atmosphere.
This effect was discussed in detail in \cite{Ambrosio4}. 
In ref.  \cite{Ambrosio5} the daily variations in solar and sidereal times were discussed .
These analyses demonstrate the high sensitivity of the detector to study very
small ($10^{-3}$ in magnitude) time variations of the muon flux.

\section{Time correlation analyses}
As already stated, it is expected that galactic CRs have a random
arrival time distribution because of the direction reshuffling of charged
CR particles by random interstellar magnetic fields. But there
may be some mechanisms which
introduce time correlations. One could expect to observe clusters in time for
events generated by charged or neutral particles coming from intermittent
emissions by nearby sources.\par

For random arrivals, the time distribution may be fitted to the Gamma function
of order M \cite{num}
\begin{equation}
G(t;\lambda,M) = N\lambda\frac{(\lambda t)^{M-1}e^{-\lambda t}}{(M-1)!}  ,
\label{eq:gamma}
\end{equation}
where,
\noindent 1/$\lambda$ is the mean value of the time difference between two
consecutive muons and N is a normalization factor.
\noindent For M=1 Eq. (1) reduces to an exponential:
\begin{equation}
G(t;\lambda,M=1) = N \lambda e^{-\lambda t}
\label{eq:gamma1}
\end{equation}
\begin{figure}[hbt]
\begin{center}
   \mbox{\epsfysize=10cm
         \epsffile{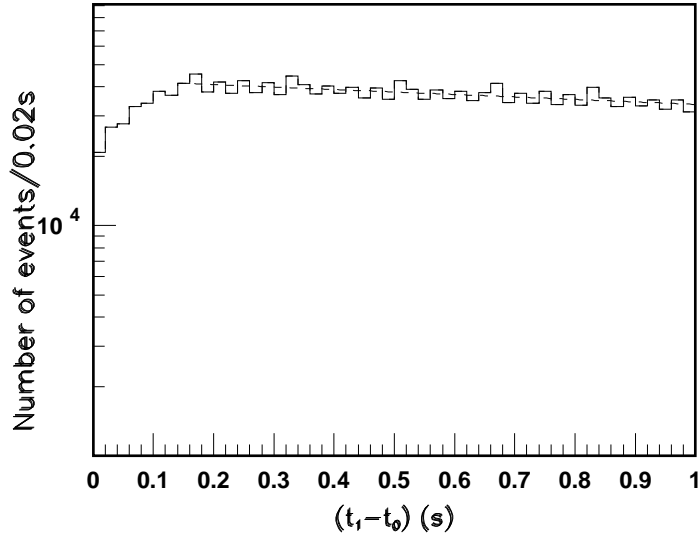}}
\end{center}
\caption{Distribution of the time between two consecutive single muons within
one second. Notice the dead time effect in the first bins.}
\label{fig:distrib1}
\end{figure}

After each trigger there was a dead time of about 100 ms (when the whole apparatus 
was operational). The effect of this dead time is evident in the (t$_1$-t$_0$)
distribution for (t$_1$-t$_0$) $<$ 100 ms, Fig. \ref{fig:distrib1}. 
In the present analysis we do not apply any correction for this dead time, but discarded 
the first point of the distributiond in the fitting procedure. 
The correction was made in a previous analysis of a sample of 0.4 million muons by
re-populating the data with events lost during the dead time \cite{Ahlen1}.

\subsection{All sky analyses }
\begin{figure}[htb]
\begin{center}
   \mbox{\epsfysize=15cm
         \epsffile{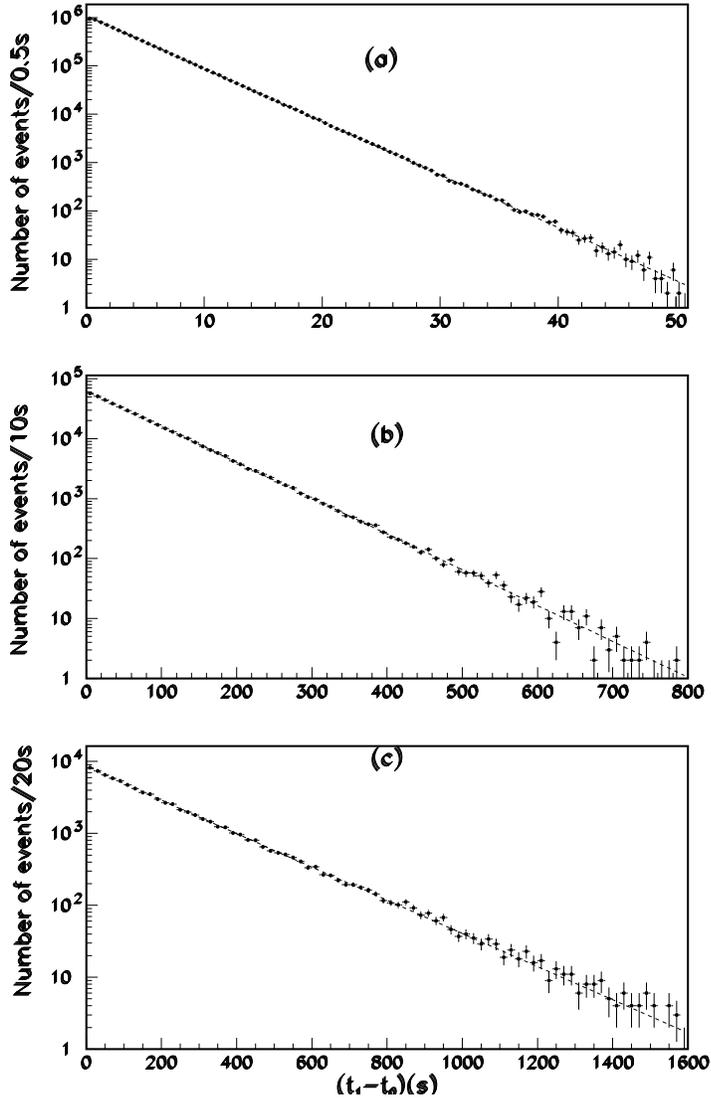}}
\end{center}
\caption{Distribution of the time between two consecutive (a)
single muons, (b) double muons and (c) multiple muons. The dashed
lines represent the fit to Eq. (\ref{eq:gamma1}). The first bin was removed
from the fits.}
\label{fig:distrib2}
\end{figure}

Fig. \ref{fig:distrib2}a shows the distribution of the time separation
between two consecutive single muons (t$_1$-t$_0$) with 0$^\circ$$\leq$zenith$\leq$72$^\circ$ 
and 0$^\circ$$\leq$azimuth$\leq$360$^\circ$. The distribution for
double  and for multiple muons are shown in Fig. \ref{fig:distrib2}b and 
 \ref{fig:distrib2}c, respectively. The three distributions are clearly 
exponential indicating the random nature of the bulk of the cosmic ray arrival
times. A fit of the data for (t$_1$-t$_0$) was made to Eq.
(\ref{eq:gamma1}): it yields the parameters quoted in Table 1.\par
\begin{figure}[htb]
\begin{center}
   \mbox{\epsfysize=15cm
         \epsffile{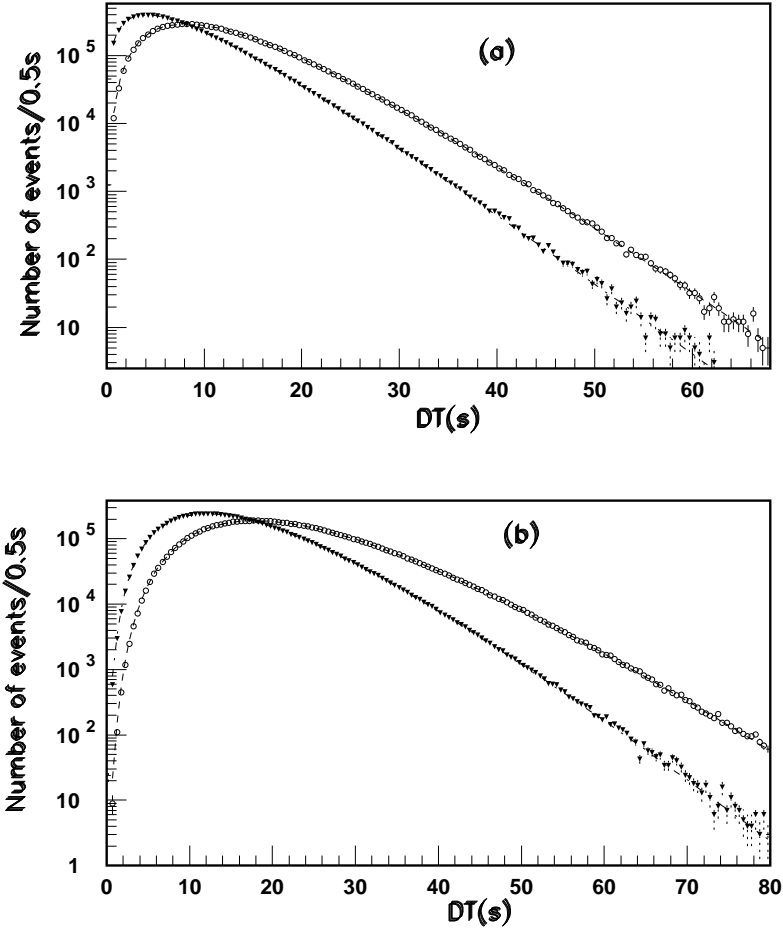}}
\end{center}
\caption{ (a) (t$_2$-t$_0$)(triangles), (t$_3$-t$_0$)(circles) and (b)
(t$_4$-t$_0$) (triangles), (t$_5$-t$_0$) (circles) time
correlations for single muons. The dashed lines represent the fits to the Gamma
functions of order 2, 3, 4 and 5, respectively.}
\label{fig:correlations1}
\end{figure}
\begin{figure}[htb]
\begin{center}				
   \mbox{\epsfysize=15cm
         \epsffile{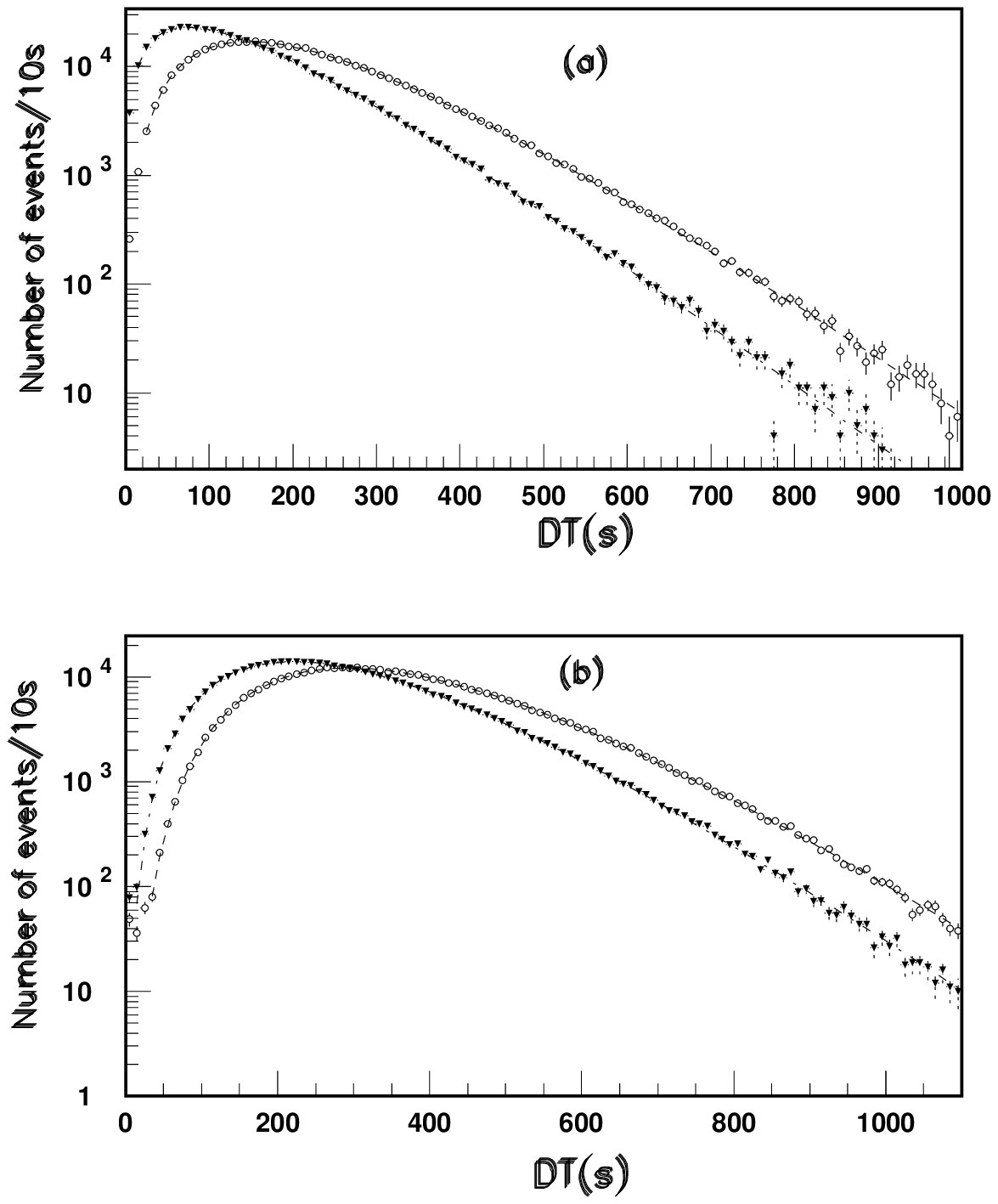}}
\end{center}
\caption{(a) the (t$_2$-t$_0$)(triangles), (t$_3$-t$_0$) (open circles) and (b)
 (t$_4$-t$_0$) (triangles), (t$_5$-t$_0$) (open circles) time
correlations for double muons. The dashed lines represent the fits to the Gamma
functions of order 2, 3, 4 and 5, respectively.}
\label{fig:correlations2}
\end{figure}

In Fig. \ref{fig:correlations1} and \ref{fig:correlations2} are presented the 
higher order correlations: (t$_2$-t$_0$), (t$_3$-t$_0$), (t$_4$-t$_0$) and 
(t$_5$-t$_0$) for single and for double muons, respectively. The experimental 
distributions are compatible
with the higher order Gamma functions, that is with random arrival
times. \par

\subsection {Analyses in narrow cones}
\begin{figure}[htb]
\begin{center}
   \mbox{\epsfysize=11cm
         \epsffile{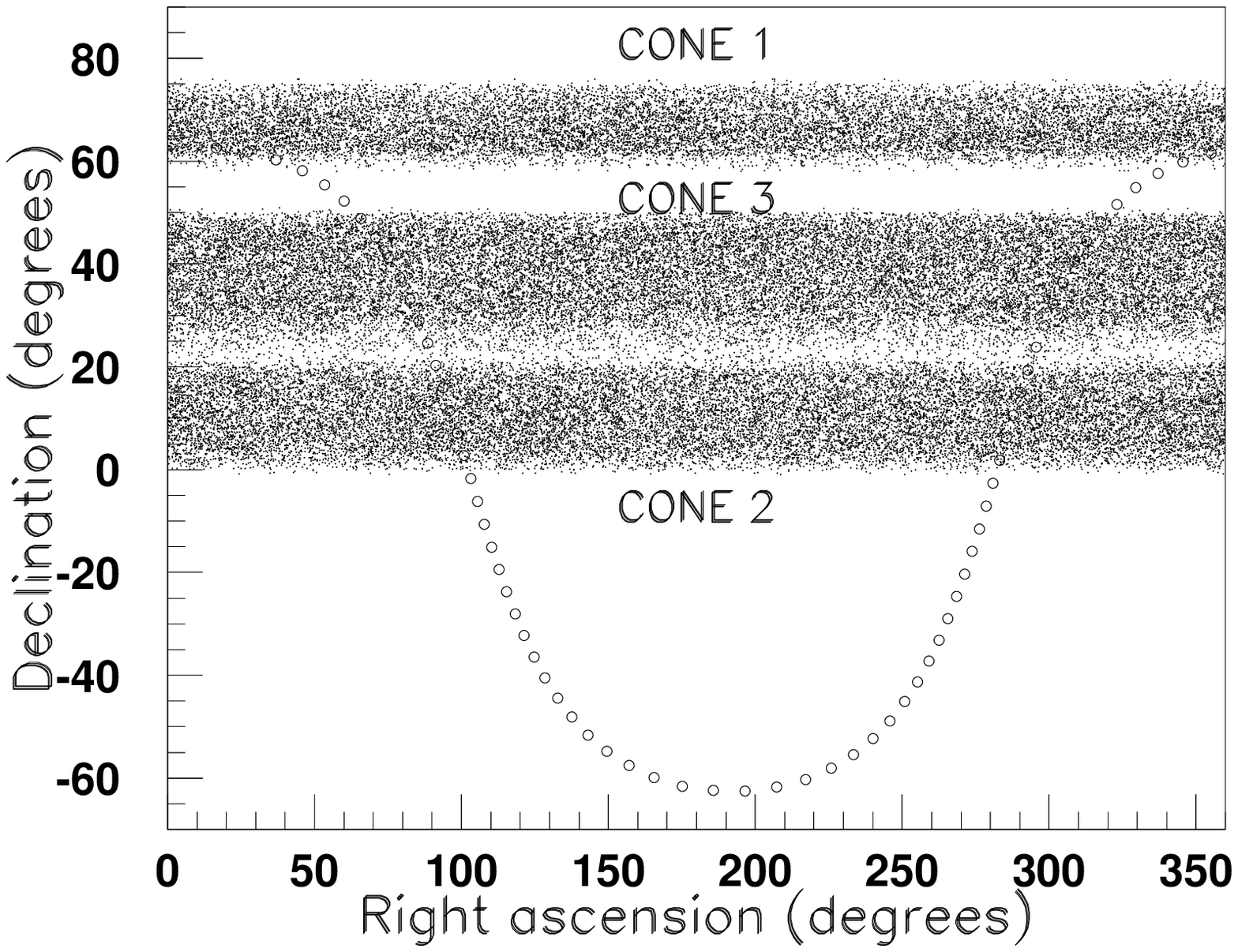}}
\end{center}
\caption{ Distributions of single muons coming from selected cones in
 declination and right ascension. The open circles indicates the galactic plane.}
\label{fig:distrib3}
\end{figure}

We repeated the same analyses on time correlations for
muons arriving from defined directions in azimuth and zenith. In this case the detector is used
as a "multiple telescope" to observe different restricted regions of the
sky, see Table 1 and Fig. \ref{fig:distrib3}. Cones (1) and (2) 
were chosen to have maximum muon intensities in the angular distributions on 
azimuth and zenith; cone (3) was chosen to cover the declination region centred 
on the direction of Cyg-X3. The pointing ability of the 
detector was checked in ref. \cite{Ambrosio6} using the Moon and 
the Sun shadows of CRs. A search for astrophysical point sources was
made using downgoing muons and none was observed, see ref. \cite{Ambrosio7}. \par 
The first cone has 25$^\circ$$\leq$zenith$\leq$45$^\circ$ and
20$^\circ$$\leq$azimuth$\leq$40$^\circ$, the total number of
selected single muons is 280$\times 10^3$. For the second cone,
25$^\circ$$\leq$zenith$\leq$45$^\circ$ and
140$^\circ$$\leq$azimuth$\leq$160$^\circ$, we selected 350$\times
10^3$ single muons. For the third cone, we have chosen
60$^\circ$$\leq$azimuth$\leq$100$^\circ$ and
260$^\circ$$\leq$azimuth$\leq$300$^\circ$ and
30$^\circ$$\leq$zenith$\leq$50$^\circ$ and the number of selected
single muons is 520$\times 10^3$. The right ascensions and declinations 
for single muons coming from the selected cones were calculated using the local coordinates 
and the times of events. In Fig. \ref{fig:distrib3}. are presented the declination bands for 
selected muons. The open circles indicates the galactic plane.\par
\begin{figure}[htb]
\begin{center}
   \mbox{\epsfysize=10cm
         \epsffile{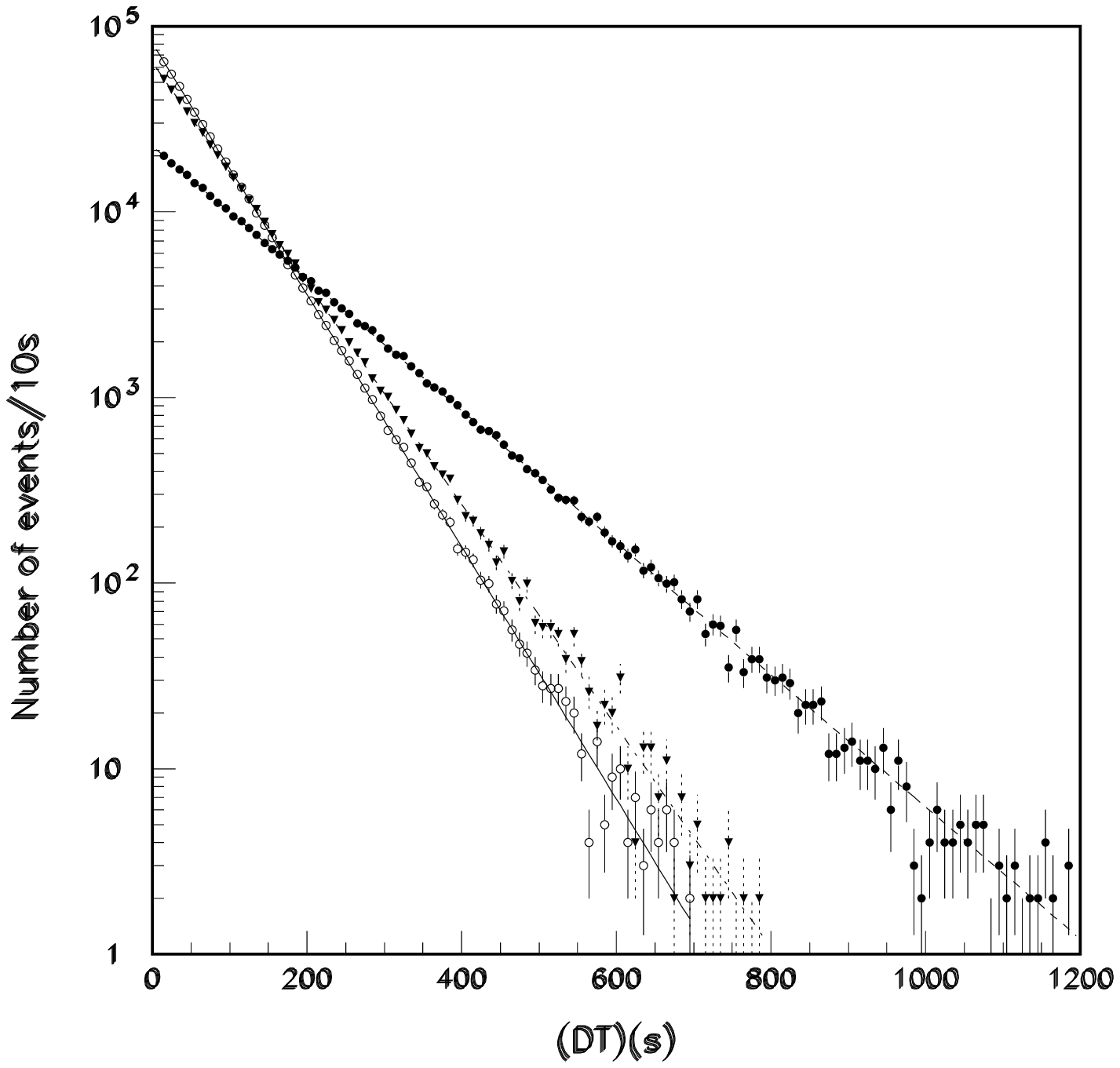}}
\end{center}
\caption{Distributions of the time between two consecutive single muons coming
from cone 1 (black circles), cone 2 (triangles) and cone 3 (open circles). The
lines represent the fits to Eq. (2).}
\label{fig:distrib4}
\end{figure}

Fig. \ref{fig:distrib4}  shows the (t$_1$-t$_0$) time distributions for single 
muons in cone 1 (black points), in cone (2) (triangles) and in cone (3) 
(open circles).

All data show the exponential character of the (t$_1$-t$_0$) distributions; 
higher order correlations agree with Eq. (\ref{eq:gamma}). It is
thus concluded that the arrival times of our selected muons are
consistent with random arrival time distributions.\par 
Table 1 gives the results of the various fits of the 
(t$_1$-t$_0$) experimental distributions to the exponential form, 
Eq. (\ref{eq:gamma1}).\\

\subsection{Kolmogorov-Smirnov test}
In order to search for possible structures in the time arrival distributions,
we have also used the Kolmogorov-Smirnov test \cite{Eadi}, which compares the
cumulative distribution $F(x)$ of the experimental data with the expected
random distribution $H(x)$. The measure of the deviation is
$d=max\mid$$F(x)-H(x)\mid$, where
$F(x)$ and $H(x)$ are the cumulative distributions of $f(x)$ (data) and
$h(x)$(expected), respectively. In terms of this quantity, $F(x)$ agrees
with $H(x,\lambda)$, where $\lambda$ is taken from the data, with a
probability of compatibility between the expected and measured distributions
given by
\begin{equation}
P_k(d > observed)=Q_{ks}(\sqrt{N}d) \label{eq:probability}
\end{equation}
where 
$$Q_{ks}(x)=2\sum_{j=1}^{+\infty} (-1)^{(j-1)}e^{-j^{2}}x^{2}$$
The probabilities of the tests for the $(t_1-t_0)$ distributions
are given in the last column of Table 1. They are consistent with random distributions, 
although, in the case of the Cygnus X3-centered cone 3, 
some disagreement (at the level of $1\sigma$) could produce the low K-S 
probability. This could originate in a possible enhancement of the event 
rate for $DT$=500 s (see Fig. 7), but the available statistics is too poor 
to reach clear conclusions. 

\begin{table}[t]
\begin{center}
\caption{Results of the fits (parameters N,$1/\lambda$, M and
$\chi^2/DoF$) of the time correlation data (t$_1$-t$_0$) to the
Gamma Function of order 1, Eq. (\ref{eq:gamma1}), for single,
double and multiple muons coming from the all sky and
from 3 narrow cones. The last column gives the probability of the
Kolmogorov-Smirnov test.}

\begin{tabular}{|c|c|c|c|c|c|c|}
\hline
Selection    &       N ($10^3$)  &  1/$\lambda$ (s) &
   M    &   $\chi^{2}$/DoF& Pr. K-S \\
\hline
Single $\mu$ &8638$\pm$4   & 4.03$\pm$0.01     &
1.002$\pm$0.003 & 1.00 & 0.99   \\

Double $\mu$&456.5$\pm$0.9     &73.3 $\pm$0.3 &
1.008$\pm$0.004 & 1.36 &0.95  \\

Multiple $\mu$   &180.2$\pm$0.3     &196.0$\pm$0.4    &
0.93$\pm$0.07& 0.98 &0.99 \\

\hline
\hline
Cone 1     &276.3 $\pm $0.6      &121.58$\pm$0.03       &
1.003$\pm$0.004  & 1.02 & 0.99\\
\hline
\hline
Cone 2     &353$\pm$1       &95.9$\pm$0.1      &
1.003$\pm$0.006  & 0.82 & 0.77 \\
\hline
\hline
Cone 3  &522$\pm$ 1 &63.82$\pm$0.3   &1.004 $\pm$0.004  &0.91&0.38   \\
\hline
\end{tabular}
\end{center}
\label{tab:table}
\end{table}

\section{Conclusions}
We have presented new high statistics data on the arrival time distributions
of downgoing cosmic ray muons with energies larger than 1.3 TeV at the top of the Gran
Sasso mountain. The data were obtained  with the streamer tube system of the
MACRO detector in its complete configuration. \par
Single, double and multiple muons arriving from the whole upper hemisphere
and also from selected space cones were considered.\par
No significant deviations from random arrival time distributions
were observed.

{\bf {Acknowledgements}}\par
\hskip 15pt
We thank the members of the MACRO Collaboration and the personnel
of the LNGS for their cooperation.
H. Dekhissi and S. Manzoor thank the Abdus Salam ICTP TRIL Programme for
providing fellowships. H. Dekhissi, J. Derkaoui, G. Giacomelli, F. Maaroufi 
and A. Moussa thank the collaboration between the Universities of Bologna 
and Mohamed Ist of Oujda.

\end{document}